\documentclass[aps,pre,groupedaddress,showpacs,twocolumn]{revtex4-1}
\usepackage{hyperref}
\usepackage{graphicx}
\usepackage{amsmath}
\usepackage{amssymb}
\usepackage{subcaption}

\begin{document}

\title{Narrow autoresonant magnetization structures in finite length
ferromagnetic nanoparticles}
\author{A. G. Shagalov$^1$}
\email{shagalov@imp.uran.ru}
\author{L. Friedland$^2$}
\email{lazar@mail.huji.ac.il}
\affiliation{$^1$Institute of Metal Physics, Ekaterinburg 620990, Russian
Federation and Ural Federal University, Mira 19, Ekaterinburg
620002, Russian Federation}
\affiliation{$^2$Racah Institute of Physics, Hebrew University of Jerusalem,
Jerusalem 91904, Israel}
\begin{abstract}
The autoresonant approach to excitation and control of large amplitude
uniformly precessing magnetization structures in finite length easy axis
ferromagnetic nanoparticles is suggested and analyzed within the
Landau-Lifshitz-Gilbert model. These structures are excited by using a
spatially uniform, oscillating, chirped frequency magnetic field, while the
localization is imposed via boundary conditions. The excitation requires the
amplitude of the driving oscillations to exceed a threshold. The dissipation
effect on the threshold is also discussed. The autoresonant driving
effectively compensates the effect of dissipation, but lowers the maximum
amplitude of the excited structures. Fully nonlinear localized autoresonant
solutions are illustrated in simulations and described via an analog of a
quasi-particle in an effective potential. The precession frequency of these
solutions is continuously locked to that of the drive, while the spatial
magnetization profile approaches the soliton limit when the length of the
nanoparticle and the amplitude of the excited solution increase.
\end{abstract}

\maketitle
\section{\label{sec:1} Introduction}

The progress in nanotechnology of magnetic materials stimulates theoretical
and experimental research of new magnetic nanostructures \cite{Braun}. It is
well known that the fundamental model of magnetization dynamics, the
Landau-Lifshitz equation, has a variety of exact solutions \cite{Kosevich}
and exhibits spatially localized objects \ -- solitons. In the small
amplitude limit, these objects were studied experimentally in thin magnetic
films \cite{Scott1,Scott2,Patton1, Patton2}. In these applications, the
solitons comprise stretched configurations and long wavelength nonlinear Schr%
\"{o}dinger (NLS) equation provides an adequate model to interpret
experimental observations \cite{Zv,Kosevich}. In nanomagnetics, the NLS
approximation for small amplitude solitons was also used in Refs. \cite%
{Li1,Li2}. In contrast, large amplitude solitons have small spatial widths
and can be observed on nanoscales only, requiring new methods of excitation
and observation. These widths can be comparable to the typical length in
magnetic materials, i.e., the width of the domain walls (usually, of order
of $10nm$). On the other hand, nanoscale samples allow to model
quasi-one-dimensional (1D) configurations \cite{Braun}, which are frequently
used in theoretical studies of solitons \cite{Kosevich} and guarantee their
stability.

In this work, we propose the autoresonant approach to excitation of large
amplitude, uniformly rotating, narrow magnetic structures in\ \textit{finite
length} easy axis ferromagnetic nanoparticles by using a weak, spatially
uniform, chirped-frequency oscillating magnetic field. The autoresonance is
a universal phenomenon used in numerous applications in many fields of
physics, e.g. in particle accelerators \cite{McMillan,Veksler}, atomic
physics \cite{Atomic1,Atomic2}, plasmas \cite{Plasma}, and nonlinear waves
\cite{L2}. A survey of mathematical problems associated with autoresonance
can be found in Ref. \cite{Kalyakin}. In recent years, autoresonance ideas
were also implemented in magnetics. Examples are autoresonant waves in
magnetic materials \cite{Sham1,Sham2}, the switching of magnetization of
single domain \textit{point-like} nanoparticles \cite{Manfredi,Lazar145},
and most recently the autoresonant excitation of large amplitude standing
magnetization waves in long ferromagnetic nanowires by using rotating
driving fields with short wavelength spatial modulations \cite{FS2019}. This
recent approach required two resonant stages, where in the first stage one
excites a rotating, but uniform magnetization of the wire, while later, in
the second resonant stage, the system develops a spatially modified standing
wave profile. In contrast, in the present work we study autoresonant
formation of localized magnetization structures in finite length
nanoparticles using a single resonance stage and without the need of a short
wavelength modulation of the driving field. The localization in this case is
imposed via boundary conditions. All these modifications require a different
theory, but the simplicity of the driving scheme is expected to facilitate
the realization of the idea in nanoscale magnets.

The paper is organized as follows. In Sec. II, we describe our model based
on the driven Landau-Lifshitz-Gilbert (LLG) equation. A weakly nonlinear
Lagrangian formulation in the dissipationless limit of this model will be
used to calculate the threshold driving amplitude for autoresonant
excitation of localized magnetic structures in Sec. III. Whitham's averaged
variational approach \cite{Whitham} will be used in the calculation. Section
IV will focus on the effect of the dissipation on the \ threshold. A fully
nonlinear, slow autoresonant dynamics of narrow chirped-driven magnetization
structures will be discussed in Sec. V and, finally, Sec. VI will summarize
our findings.

\section{\label{sec:2}Autoresonant excitation model}

We consider a quasi-one-dimensional ferromagnetic nanoparticle of length $L$
oriented along the $z$-axis. The particle has an easy-axis anisotropy along $%
z$ and is located in a constant external magnetic field $\mathbf{H}=H_{0}%
\widehat{\mathbf{e}_{z}}$ combined with a weak uniform rotating driving
field $\mathbf{H}_{d}(t)=g(\cos \varphi _{d}\widehat{\mathbf{e}}_{x}+\sin
\varphi _{d}\widehat{\mathbf{e}}_{y})$ having slowly chirped rotation
frequency $\omega _{d}(t)=-\partial \varphi _{d}/\partial t$. We model this
system by LLG equation (in dimensionless form):
\begin{equation}
\mathbf{m}_{\tau }=\mathbf{h\times m+}\eta \mathbf{m\times m}_{\tau },
\label{L1}
\end{equation}%
were $\mathbf{m=M/}M$ is the normalized magnetization, $\eta $ is the
damping parameter, and%
\begin{equation}
\mathbf{h}=\mathbf{m}_{\xi \xi }+(m_{z}+h_{0})\widehat{\mathbf{e}}%
_{z}+\varepsilon (\cos \varphi _{d}\widehat{\mathbf{e}}_{x}+\sin \varphi _{d}%
\widehat{\mathbf{e}}_{y}).  \label{L2}
\end{equation}%
Here and throughout the rest of the text $(...)_{\tau }$ and $(...)_{\xi }$
denote partial derivatives with respect to $\tau $ and $\xi $. We use
dimensionless time $\tau =(\gamma K/M)t$ and coordinate $\xi =z/\delta $, $%
(\delta =\sqrt{A/K})$, where $\gamma $, $A$, and $K$ are the gyromagnetic
ratio, the exchange constant, and the anisotropy constant, respectively. In
Eq. (\ref{L2}), $h_{0}=MH_{0}/K$, $\varepsilon =Mg/K$, $\varphi _{d}=-\int
\Omega _{d}d\tau $, $\Omega _{d}(\tau )=\omega _{d}M/(K\gamma )$. We also
assume that initially the magnetization $m_{z}(\xi ,0)=1$ is uniform, while
at the ends $\xi =0$ and $\xi =l=L/\delta $ of the particle remains fixed ($%
m_{z}=1$) at all times. An approach to realization of these boundary
conditions will be discussed in Sec. V.

As in many other driven nonlinear systems \cite{Lazar125}, the
autoresonant excitation of rotating localized magnetization structures
requires (a) slow passage of the driving frequency $\Omega _{d}(\tau )$
through a resonant frequency $\Omega _{0}$, in our case, $\Omega
_{0}=1+k^{2}+h_{0}$ (here $k=\pi /l$), and (b) that the driving amplitude $%
\varepsilon $ exceeds some threshold value $\varepsilon _{th}$. In
particular, in the simplest case $\Omega _{d}(\tau )=\Omega _{0}-\alpha \tau
$, we expect the characteristic scaling $\varepsilon _{th}\sim \alpha ^{3/4}$
\cite{Lazar125}, where $\alpha $ is the driving frequency chirp rate. When
these two conditions are met, the driven magnetic perturbation will be
captured into a continuing nonlinear resonance (phase-locking) with the
drive, leading to large amplitude excitations of the magnetization, as the
frequency chirp continues. An example of a system under consideration could
be a Permalloy sample with $A=10^{-11}J/m$, $K=10^{5}J/m^{3}$, and $%
M=8\times 10^{5}A/m$ \cite{W}. In this case, the characteristic magnetic
length is $\delta \approx 10nm$ and the exchange length $\delta _{m}=2\sqrt{%
A/\mu _{0}M_{0}^{2}}\approx 7nm$. It is well known that the value $\pi
\delta $ represents a typical width of domain walls. The solitons in
easy-axis magnets \cite{Kosevich} can be interpreted as two interacting
domain walls of opposite topological signs (a "breather" in terms of Ref.
\cite{Braun}) and the soliton width can be estimated as $2\pi \delta $. A
simplest nanoparticle, where large amplitude narrow magnetic structures\
having near soliton spatial profiles can be observed is a segment of a
ferromagnetic nanowire of length $L>2\pi \delta $ and cross-section $d<2\pi
\delta _{m}$. The small cross-section guarantees quasi-one-dimensionality of
the system \cite{Braun} in the direction of the segment, which is assumed in
our model [Eqs. (\ref{L1}) and (\ref{L2})].

\section{\label{sec:3}The threshold phenomenon}

We proceed by illustrating the autoresonant excitation and the threshold
phenomenon in system (\ref{L1}) in numerical simulations. We used the
numerical approach described in Ref. \cite{FS2019}. The simulations solved
an equivalent system of two coupled NLS-type equations based on the quantum
two-level analog due to R. Feynman \cite{Feynman}. The numerical scheme used
a standard pseudospectral method \cite{Canuto} subject to given initial and
boundary conditions, $m_{z}(\xi ,0)=1$ and $m_{z}(0,\tau )=m_{z}(1,\tau )=1$%
. In the following illustration, we neglected dissipation and used the
driving frequency $\Omega _{d}(\tau )=h_{0}+\Omega _{d}^{\prime }(\tau )$,
where
\begin{equation}
\Omega _{d}^{\prime }=\left\{
\begin{array}{c}
1+k^{2}-\Delta \omega \sin (\alpha \tau /\Delta \omega ),\text{ \ }\tau <\pi
\Delta \omega /(2\alpha ), \\
1+k^{2}-\Delta \omega ,\text{ \ }\tau >\pi \Delta \omega /(2\alpha ),%
\end{array}%
\right.  \label{S3_om}
\end{equation}%
yielding a quasi-steady-state solution in the final stage of excitation, as $%
\Omega _{d}^{\prime }(\tau )$ approached a constant. The parameters were $%
\alpha =0.005$, $l=8$, and $\Delta \omega =0.9$. Figure 1 shows $%
-m_{z}=-\cos \theta $ and phase mismatch $\Phi =\Phi =\varphi -\varphi _{d}$
($\theta $ and $\varphi $ being the spherical coordinates of the
magnetization vector) versus slow time $T=\sqrt{\alpha }\tau $ just above ($%
\varepsilon =1.05\varepsilon _{th}^{0}$) and below ($\varepsilon
=0.95\varepsilon _{th}^{0}$) the threshold $\varepsilon _{th}^{0}=0.01$ [see
Eq. (\ref{S3_11})]. Panels a,b in the figure demonstrate the excitation of a
localized large amplitude magnetic structure, which is phase-locked to the
drive. In contrast, in panels c,d for $\varepsilon <\varepsilon _{th}^{0}$,
phase-locking is destroyed and the excitation saturates at some small
amplitude.

\begin{figure}[bp]
\begin{subfigure}[bp]{0.48\textwidth}
    \includegraphics[width=\textwidth]{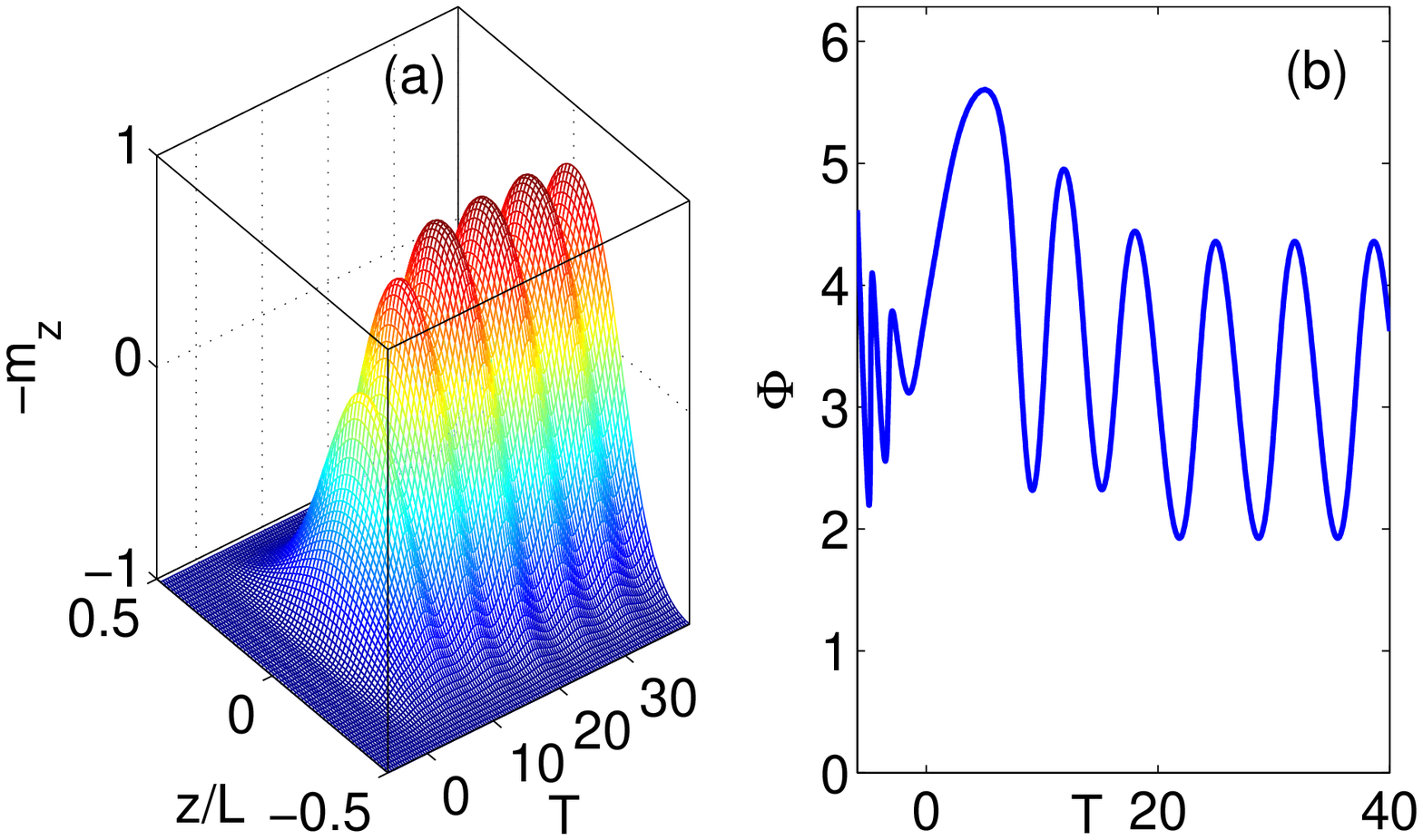}
  \end{subfigure}
\begin{subfigure}[bp]{0.48\textwidth}
    \includegraphics[width=\textwidth]{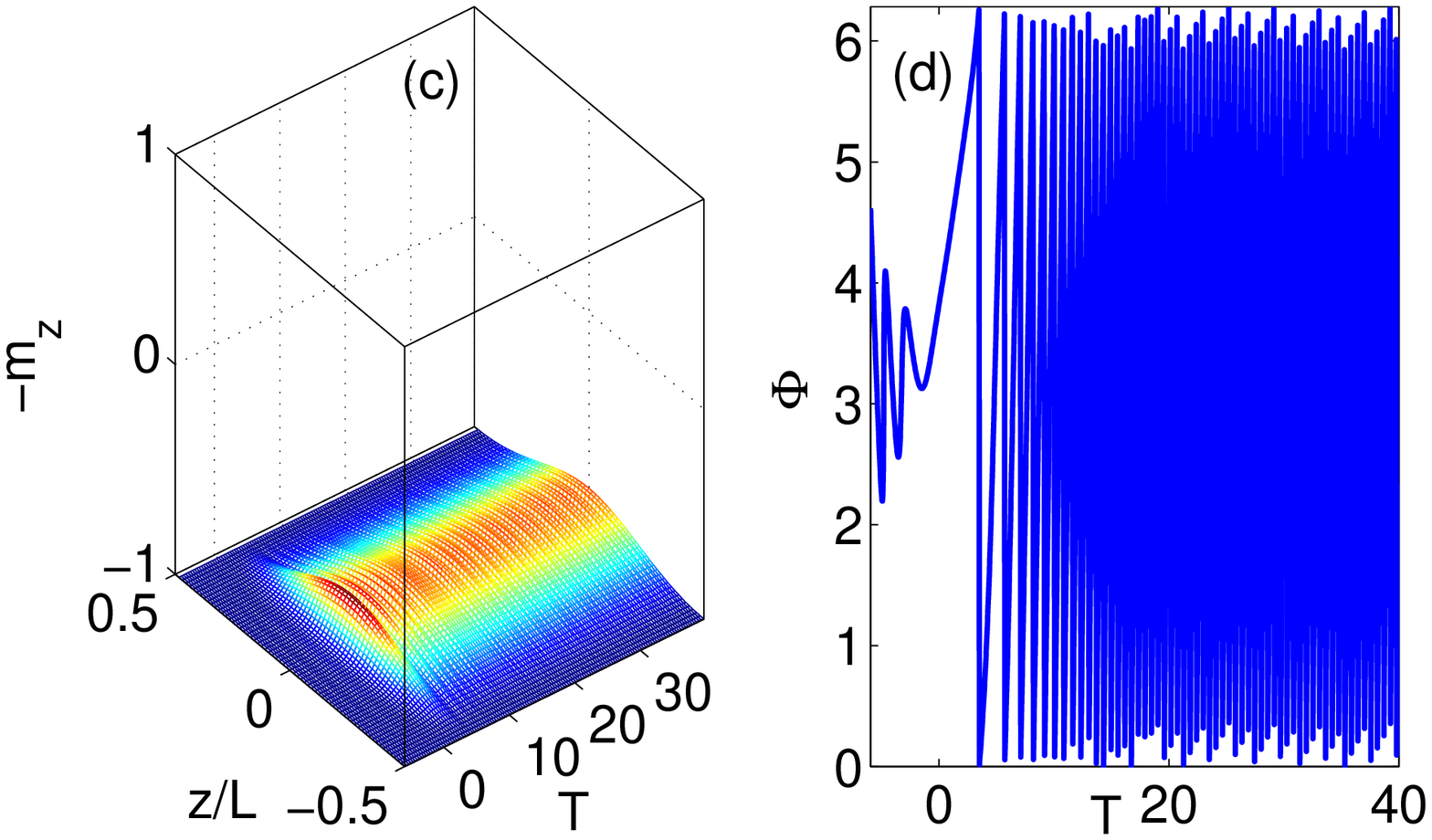}
  \end{subfigure}
\caption{The magnetization component $-m_{z}$ and phase mismatch $\Phi $
versus slow time $T=\protect\alpha ^{1/2}t$. (a) and (b) just above the
threshold, $\protect\varepsilon =1.05\protect\varepsilon _{th}^{0}$. (c) and
(d) just below the threshold, $\protect\varepsilon =0.95\protect\varepsilon %
_{th}^{0}$. }
\label{Fig1}
\end{figure}
In analyzing the autoresonance threshold in the problem analytically, we
first use the dissipationless limit of Eqs. (\ref{L1}) in spherical
coordinates: $m_{x}=\sin \theta \cos \varphi $, $m_{y}=\sin \theta \sin
\varphi $, $m_{z}=\cos \theta $:%
\begin{eqnarray}
\theta _{\tau } &=&\Phi _{\xi \xi }\sin \theta +2\Phi _{\xi }\theta _{\xi
}\cos \theta -\varepsilon \sin \Phi ,  \label{S3_1} \\
\Phi _{\tau } &=&-\frac{\theta _{\xi \xi }}{\sin \theta }+\Phi _{\xi
}^{2}\cos \theta +\cos \theta -\Omega _{d}^{\prime }-\varepsilon \cot \theta
\cos \Phi .  \label{S3_2a}
\end{eqnarray}%
The boundary conditions are $\theta (0)=\theta (l)=0$. Equations (\ref{S3_1}%
), (\ref{S3_2a}) satisfy the variational principle with the Lagrangian
density
\begin{eqnarray}
\Lambda  &=&\frac{1}{2}(\theta _{\xi }^{2}+\Phi _{\xi }^{2}\sin ^{2}\theta
)+\Phi _{\tau }\cos \theta   \label{S3_3} \\
&&+\Omega _{d}^{\prime }(\tau )\cos \theta -\frac{1}{4}\cos (2\theta
)-\varepsilon \sin \theta \cos \Phi ,  \notag
\end{eqnarray}%
which allows to write Eqs. (\ref{S3_1}), (\ref{S3_2a}) as
\begin{eqnarray}
\left( \frac{\partial \Lambda }{\partial \Phi _{\tau }}\right) _{\tau
}+\left( \frac{\partial \Lambda }{\partial \Phi _{\xi }}\right) _{\xi } &=&%
\frac{\partial \Lambda }{\partial \Phi },  \label{S3_3a} \\
\left( \frac{\partial \Lambda }{\partial \theta _{\xi }}\right) _{\xi } &=&%
\frac{\partial \Lambda }{\partial \theta }.  \label{S3_3b}
\end{eqnarray}%
Since the autoresonance threshold is a weakly nonlinear phenomenon \cite%
{Lazar145}, our next goal is to discuss the slow, weakly nonlinear evolution
in the problem and we use Whitham's averaged Lagrangian approach \cite%
{Whitham} for achieving this goal. This approach is designed to describe
wave systems with slow parameters. In our case the \textit{slow} parameter
is the driving frequency $\Omega _{d}(\tau )$ and the slowness means that
this frequency does not change significantly during one period of the
driving, i.e. $\frac{2\pi }{\Omega _{d}^{2}}\frac{d\Omega _{d}}{d\tau }\ll 1.
$ We seek \textit{rapidly} rotating solution, such that $\varphi
_{t}\approx \Omega _{d}(\tau )$, but assume that the phase mismatch $\Phi
(\tau )=\varphi -\varphi _{d}$ is slow (i.e., experiences only a small
change during one period of the drive). Since in the linear approximation
and constant $\Omega _{d}$, Eqs. (\ref{S3_1}), (\ref{S3_2a}) yield a
phase-locked driven solution
\begin{equation*}
\theta =a\sin (k\xi ),\Phi =\pi
\end{equation*}%
where $a=\varepsilon /(\Omega _{d}^{\prime }-1-k^{2})$, in the \textit{%
slowly varying }$\Omega _{d}$ problem, we use the following \textit{small}
amplitude ansatz
\begin{equation}
\theta =a(\tau )\sin \Theta ,\Phi =f(\tau )+b(\tau )\sin \Theta .
\label{S3_4}
\end{equation}%
Here $\Theta =k\xi $ and it is assumed that the weak nonlinearity, small
driving, and slow variation of the driving frequency introduce slow time
dependence in $a$, $b$ and $f$. Note that this anzatz is just a truncated
Fourier expansion conserving the boundary conditions and that for $k=\pi /l$%
, the length $l~$of the nanoparticle is one half of the periodicity length $%
l_{0}=2\pi /k$ in (\ref{S3_4}). At this stage, for simplifying the
derivation, we will set $b=0$, but later show that to lowest order,
inclusion of nonzero $b$ does not introduce a significant change in the
weakly nonlinear theory. The substitution of (\ref{S3_4}) into (\ref{S3_3}),
the expansion to fourth order in $a$, and averaging over $\Theta \in \lbrack
0,\pi ]$ yields the averaged Lagrangian density%
\begin{equation}
\bar{\Lambda}=\frac{a^{2}}{4}(1+k^{2}-\Omega _{d}^{\prime }-\beta )+\frac{%
a^{4}}{64}(-4+\Omega _{d}^{\prime }+\beta )-\frac{2}{\pi }a\varepsilon \cos
f,  \label{S3_5}
\end{equation}%
where $\beta =f_{\tau }$ and we omitted terms independent of $\ f$ and $a$,
as not contributing the dynamics. Note that Eqs. (\ref{S3_4}) involve only
slow time dependencies and, thus, there is no need for averaging the
Lagrangian over the fast oscillations. Next, we use $\bar{\Lambda}$ and take
variation with respect to\ $f$, leaving lowest significant order terms only
to get%
\begin{equation}
\frac{da}{d\tau }=-\frac{4\varepsilon }{\pi }\sin f.  \label{S3_6}
\end{equation}%
Here and in the next section the problem depends on $\tau $ only, so the
time derivatives are now defined as $d(...)/dt$. Similarly, the variation
with respect to $a$ yields the second evolution equation%
\begin{equation}
\beta =\frac{df}{d\tau }=1+k^{2}-\Omega _{d}^{\prime }-\frac{4-\Omega
_{d}^{\prime }}{8}a^{2}-\frac{4\varepsilon }{\pi a}\cos f.  \label{S3_7}
\end{equation}%
Note that the frequency of rotation of the magnetization vector in the
linearized undriven problem is $h_{0}+1+k^{2}$ and assume passage through
the linear resonance, i.e. $\Omega _{d}^{\prime }=\Omega
_{d}-h_{0}=1+k^{2}-\alpha \tau $, where $\alpha \tau $ is viewed as a small
deviation \ of the driving frequency from the resonance. Then Eqs. (\ref%
{S3_6}) and (\ref{S3_7}) guarantee the assumed slowness of variation of $a$
and $f$ if the nonlinearity and driving amplitude $\varepsilon $ and chirp
rate $\alpha $ are sufficiently small. Next, we introduce rescaled amplitude
$A=[(3-k^{2})/8]^{1/2}\alpha ^{-1/4}a$. This allows to rewrite Eqs. (\ref%
{S3_6}) and (\ref{S3_7}) as
\begin{eqnarray}
\frac{dA}{dT} &=&-\mu \sin f,  \label{S3_8} \\
\frac{df}{dT} &=&T-A^{2}-\frac{\mu }{A}\cos f,  \label{S3_9}
\end{eqnarray}%
where $T=\alpha ^{1/2}\tau $ is the slow time (used in Figs. 1 and 2), and
\begin{equation}
\mu =\frac{[2(3-k^{2})]^{1/2}\varepsilon }{\pi \alpha ^{3/4}}.  \label{S3_10}
\end{equation}%
By introducing a complex dependent variable $\Psi =Ae^{if}$, Eqs. (\ref{S3_8}%
) and (\ref{S3_9}) can be combined into a single equation characteristic of
many autoresonance problems \cite{Lazar125}:%
\begin{equation}
i\frac{d\Psi }{dT}+(T-|\Psi |^{2})\Psi =\mu .  \label{S3_10a}
\end{equation}%
If starting in zero equilibrium $\Psi =0$ at $T<0$ (above the linear
resonance), this equation guarantees phase locking at $f\approx \pi $ after
passage through the resonance (i.e., for $T=0$), where $A$ increases as $%
\sim \sqrt{T}$ , provided the single parameter $\mu $ in the problem exceeds
the value of $0.41$ \cite{Lazar145}. This yields the threshold driving
amplitude
\begin{equation}
\varepsilon _{th}^{0}=\frac{0.41\pi \alpha ^{3/4}}{[2(3-k^{2})]^{1/2}}.
\label{S3_11}
\end{equation}%
The results in Fig. 1 illustrate that this threshold is in a good agreement
with numerical simulations.

Next, we include the spatial variation in $\Phi =f(\tau )+b(\tau )\sin
\Theta $ in calculating the averaged Lagrangian density. This results in the
addition of a new, $b$-dependent part
\begin{equation}
\bar{\Lambda}_{b}(a,f,b,b_{\tau })=\frac{2}{3\pi }a^{2}b_{\tau }+\frac{%
\varepsilon }{2}ab\sin f+\frac{k^{2}}{16}a^{2}b^{2}  \label{S3_12}
\end{equation}%
to already discussed Lagrangian density (\ref{S3_5}). By taking the
variation with respect to $b$, we now get a new evolution equation%
\begin{equation}
\frac{da}{d\tau }=-\frac{3\pi }{8}\varepsilon \sin f-\frac{3\pi k^{2}}{32}ab.
\label{16}
\end{equation}%
Then, on \ using Eq. (\ref{S3_6}), one obtains
\begin{equation}
ab=\frac{32}{3\pi k^{2}}(\frac{4}{\pi }-\frac{3\pi }{8})\varepsilon \sin f=%
\frac{0.32}{k^{2}}\varepsilon \sin f.  \label{17}
\end{equation}%
This result shows that $b$ becomes negligibly small as $a$ increases and
significantly exceeds $\varepsilon $ in the autoresonant stage, validating
the derivation of the threshold driving amplitude presented above. This
effect is illustrated in Fig. 2, showing the spatial form of $\Phi $ versus
slow time $T$ for parameters $\varepsilon =0.02$, $\alpha =0.005$, $L=8$,
and $\Delta \omega =1$. One can see how the spatial modulation of $\Phi $
nearly disappears after the passage through the linear resonance at $T=0$
and entering the autoresonant stage of evolution.
\begin{figure}[tp]
\centering\includegraphics[width=8.8cm]{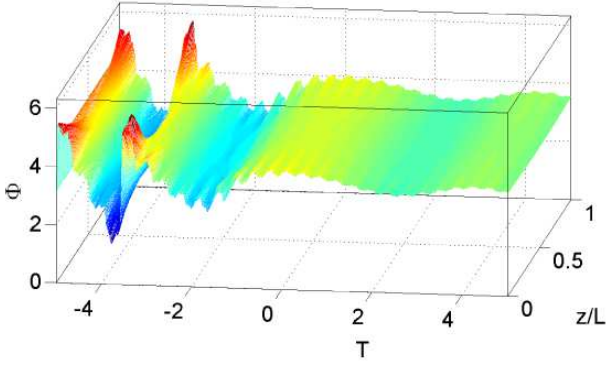}
\caption{The disappearance of the spatial modulation of $\Phi $ after
entering the autoresonance regime}
\label{Fig2}
\end{figure}

%\section{\label{sec:4} Effect of dissipation}

\section{Effect of dissipation}

For including a weak dissipation in the calculation of the threshold, we
return to the original system of evolution equations with Gilbert damping

\begin{eqnarray}
\left( \frac{\partial \Lambda }{\partial \Phi _{\tau }}\right) _{\tau
}+\left( \frac{\partial \Lambda }{\partial \Phi _{\xi }}\right) _{\xi }-%
\frac{\partial \Lambda }{\partial \Phi } &=&-\eta \varphi _{\tau }\sin
^{2}\theta ,  \label{S4_1} \\
\left( \frac{\partial \Lambda }{\partial \theta _{\xi }}\right) _{\xi }-%
\frac{\partial \Lambda }{\partial \theta } &=&\eta \theta _{\tau }.
\label{S4_2}
\end{eqnarray}%
\begin{figure}[tp]
\centering\includegraphics[width=8.8cm]{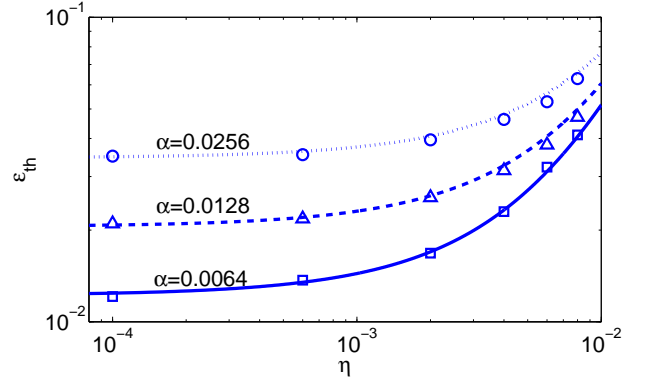}
\caption{The effect of the dissipation on the threshold for capture into
autoresonance. The lines are given by Eq. (\protect\ref{S4_7}) and the
markers represent numerical simulations.}
\label{Fig3}
\end{figure}
Here the left hand sides are the Lagrange's components of the
dissipationless case described by the Lagrangian density (\ref{S3_3}) and,
as before, $\varphi =\Phi +\varphi _{d}$ is the azimuthal angle of the
magnetization vector. Within the ansatz $\Phi =f(\tau )$ and $\theta
=a(t)\sin \Theta $ discussed above, Eq. (\ref{S4_1})\ yields
\begin{equation}
\theta _{\tau }\sin \theta +\varepsilon \sin \theta \sin \Phi =\eta (\Phi
_{\tau }-\Omega _{d})\sin ^{2}\theta ,  \label{S4_3}
\end{equation}%
where $\Omega _{d}=d\varphi _{d}/d\tau =\Omega _{0}-\alpha \tau $ is the
driving frequency. On averaging the last equation with respect to $\Theta
\in \lbrack 0,\pi ],$ one obtains%
\begin{equation}
\frac{da}{d\tau }=-\frac{4\varepsilon }{\pi }\sin f-\eta a\Omega _{d},
\label{S4_4a}
\end{equation}%
where, assuming a continuing phase-locking ($\Phi \ $remains nearly
constant), we neglected $\Phi _{\tau }$ in the right hand side. The last
equation replaces Eq. (\ref{S3_6}) of the dissipationless case. Furthermore,
this equation shows that the right hand side of Eq. (\ref{S4_2}) involves a
product of two small objects and can be neglected in the following. Thus,
the additional evolution equation [originating from (\ref{S4_2})]
\begin{equation}
\frac{df}{d\tau }=\alpha \tau -\frac{3-k^{2}}{8}a^{2}-\frac{4\varepsilon }{%
\pi a}\cos f,  \label{S4_5}
\end{equation}%
is the same as Eq. (\ref{S3_7}) for the dissipationless case.

Using the rescaling of the dependent and independent variables of the
dissipationless case, Eqs. (\ref{S4_4a}) and (\ref{S4_5}) can be combined
into a single complex equation [compare to Eq. (\ref{S3_10a})]%
\begin{equation}
i\frac{d\Psi }{dT}+(T-|\Psi |^{2})\Psi +i\frac{\gamma }{2}\Psi =\mu ,
\label{S4_6}
\end{equation}%
where $\gamma =2\eta \Omega _{d}\alpha ^{-1/2}$. The capture into
autoresonance in this problem was studied in Ref. \cite{Lazar121}, yielding
the following threshold driving amplitude for $\gamma <1$.
\begin{equation}
\varepsilon _{th}\approx \varepsilon _{th}^{0}(1+1.06\gamma +0.67\gamma
^{2}).  \label{S4_7}
\end{equation}%
We compare this result with the numerical simulations of the original LLG
equation in Fig. 3, showing $\varepsilon _{th}$ versus $\eta $ for three
values of the chirp rate $\alpha =0.0064$, $0.0128$, and $0.0256$. We used
parameters $l=8$, $\Delta \omega =0.9$, $h_{0}=5$ and frequency variation of
the form (\ref{S3_om}). The simulation was limited to values of $\gamma <1$
and used the initial and final simulation times $T_{0}=-10$ and $T_{1}=30$,
respectively, reaching the quasi-steady-state at the final time. Arriving in
simulations at this quasi-steady-state without dephasing served as the
criterion for finding $\varepsilon _{th}$.

\section{Nonlinear quasi-steady-state}

In our numerical simulations (e.g., Fig.1) we observe that in the
autoresonant stage of evolution, both $\Phi $ and $\theta $ perform slow
oscillations around some smooth time varying averages. We will refer to the
smooth evolution as the quasi-steady-state (it still slowly varies in time).
Similar slow oscillations around the smooth average were observed and
studied in many autoresonant problems \cite{Lazar125} and reflect the
stability of the quasi-steady-state if perturbed. The problem of stability
of fully nonlinear autoresonant evolution is important, but very complex in
our case and will remain outside the scope of the present work and, thus, we
focus on describing the quasi-steady-state only. We proceed from Eqs. (\ref%
{S3_1}) and (\ref{S3_2a}) and assume a perfect continuing phase-locking, $%
\Phi =\pi $ (uniform precession of the magnetization vector, $\varphi
_{t}=\Omega _{d}^{\prime }(\tau )$), and neglect $\theta _{\tau }$ despite
the time variation of the driving frequency. This leaves us with a single
equation%
\begin{equation}
\theta _{\xi \xi }=-V_{\theta },  \label{S5_1}
\end{equation}%
where we have also neglected the driving term and defined the effective
potential
\begin{equation}
V=-\Omega _{d}^{\prime }(\tau )\cos \theta +\frac{1}{4}\cos (2\theta )-[%
\frac{1}{4}-\Omega _{d}^{\prime }(\tau )].  \label{S5_2}
\end{equation}%
We have added a shift in $V$ so that $V=0$ at $\theta =0$ at all times,
which is convenient for classifying different types of solutions (see
below). Equation (\ref{S5_1}) can be viewed as defining the motion of a
quasiparticle having coordinate $\theta $ in the effective potential $V$, $%
\xi $ playing the role of time and $\Omega _{d}^{\prime }(\tau )~$being a
parameter. The solution of (\ref{S5_1}) subject to boundary conditions $%
\theta =0$ ($m_{z}=1)$ at $\xi =0,l$, corresponds to a \textit{single} round
trip of the quasiparticle in the effective potential starting from $\theta
=0 $ at $\xi =0$ and returning to the same point $\theta =0$ at $\xi =l$.

\begin{figure}[tp]
\centering\includegraphics[width=8.8cm]{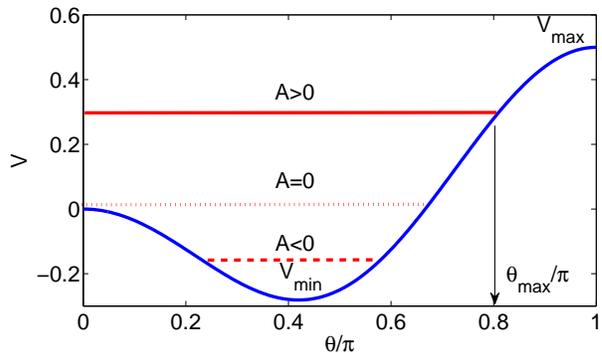}
\caption{The effective potential. Two classes of oscillations correspond to
different values of the quasi-energy: $A>0$ (solid horizontal line)is the
case of the zero boundary conditions used in this work, $A<0$ (dished
horizontal line)is the standing magnetization wave of Ref. \protect\cite%
{FS2019}. The limit $A\rightarrow 0$ (dotted horizontal line) corresponds to
the soliton solution of infinite extent ($\protect\lambda \rightarrow \infty
$).}
\label{Fig4}
\end{figure}
The form of the quasi-potential allows to qualitatively describe the motion
of the quasiparticle in the problem. Note that for $\Omega _{d}^{\prime }<1$
the quasipotential has a zero maximum at $\theta =0$, a minimum $V_{\min }=-$
($\Omega _{d}^{\prime 2}/2+1/4)$ at $\cos \theta =$ $\Omega _{d}^{\prime }$,
and another maximum $V_{\max }=2\Omega _{d}^{\prime }$ at $\theta =\pi $, as
illustrated in Fig. 4 for $\Omega _{d}^{\prime }=0.25$. Therefore, the
quasiparticle passes $\theta =0$ (i.e. one satisfies the boundary conditions
$\theta =0$ at $\xi =0$ and $l$) provided (a) $V_{\max }>0$ (i.e. the
allowed range of $\Omega _{d}^{\prime }$ is $0<\Omega _{d}^{\prime }<1$) and
(b) the quasi-energy $A(\tau )=\frac{1}{2}\theta _{\xi }^{2}+V$ \ is in the
interval $[0,V_{\max }]$ (see the solid red horizontal line in Fig. 4). The
solutions for $\theta (\xi )$ in this case have a single maximum and a
finite extent in $\xi $. In addition to this type of solutions, there also
exists a motion of the quasiparticle, such that $V_{\min }<A<0$ (see the
horizontal red dashed line in Fig.4), but this motion does not reach $\theta
=0$ and, thus, can not satisfy our boundary conditions. Autoresonant
magnetic excitations of this different type were discussed in Ref. \cite%
{FS2019} and comprise spatial oscillations of $\theta $  having a finite
periodicity length. Interestingly, the two types of solutions coalesce in the
limit $A\rightarrow 0$ (the dotted red line in Fig. 4) and $\theta (\xi )$
assumes a form of a single peak defined on the infinite $\xi $ domain, i.e a
soliton (see Ref \cite{Kosevich}). The physical width of $\theta $ in this
limit can be estimated as
\begin{equation}
\Delta \xi \approx 2\pi /\kappa _{m},  \label{Width}
\end{equation}%
where $\kappa _{m}=(1-\Omega _{d}^{\prime 2})^{1/2}$ is the frequency of
linear oscillations around the minimum of $V$. For example, if $0<\Omega
_{d}^{\prime }<0.5$, one has $2\pi <\Delta \xi <7.25$. In our case of a
finite length of the nanoparticle, $\theta (\xi )$ can only approximately
approach the form of the soliton, but, nevertheless, Eq. (\ref{Width}) can
still be used as an estimate of the with of the excited solution.

Note that Eq. (\ref{S5_1}) can be solved in quadratures for finding $\theta
(\xi ,\tau )$ at a given time. However, this solution also requires the
knowledge of the quasi-energy $A(\tau )=\frac{1}{2}\theta _{\xi }^{2}+V$.
Finding this energy, seems to require including the drive and solving the
full driven PDE. Nevertheless, if the system evolves in autoresonance, one
can calculate $A(\tau )$ by using simplified arguments. Indeed, in the
autoresonance, the azimuthal frequency of magnetization follows that of the
drive, while the extent $\lambda $ of the spatial profile of $\theta $
remains constant $\lambda =l$. The preservation of $\lambda (\tau )$ at
value $l$, despite the variation of $\Omega _{d}^{\prime }$ in time allows
calculation of the quasi-energy at all times using the definition
\begin{equation}
\lambda =2\int_{0}^{\theta _{\max }}\frac{d\theta }{\sqrt{2[A(\tau
)-V(\theta ,\Omega _{d}^{\prime })]}}.  \label{Lambda}
\end{equation}%
\begin{figure}[bp]
\begin{subfigure}[bp]{0.48\textwidth}
    \includegraphics[width=\textwidth]{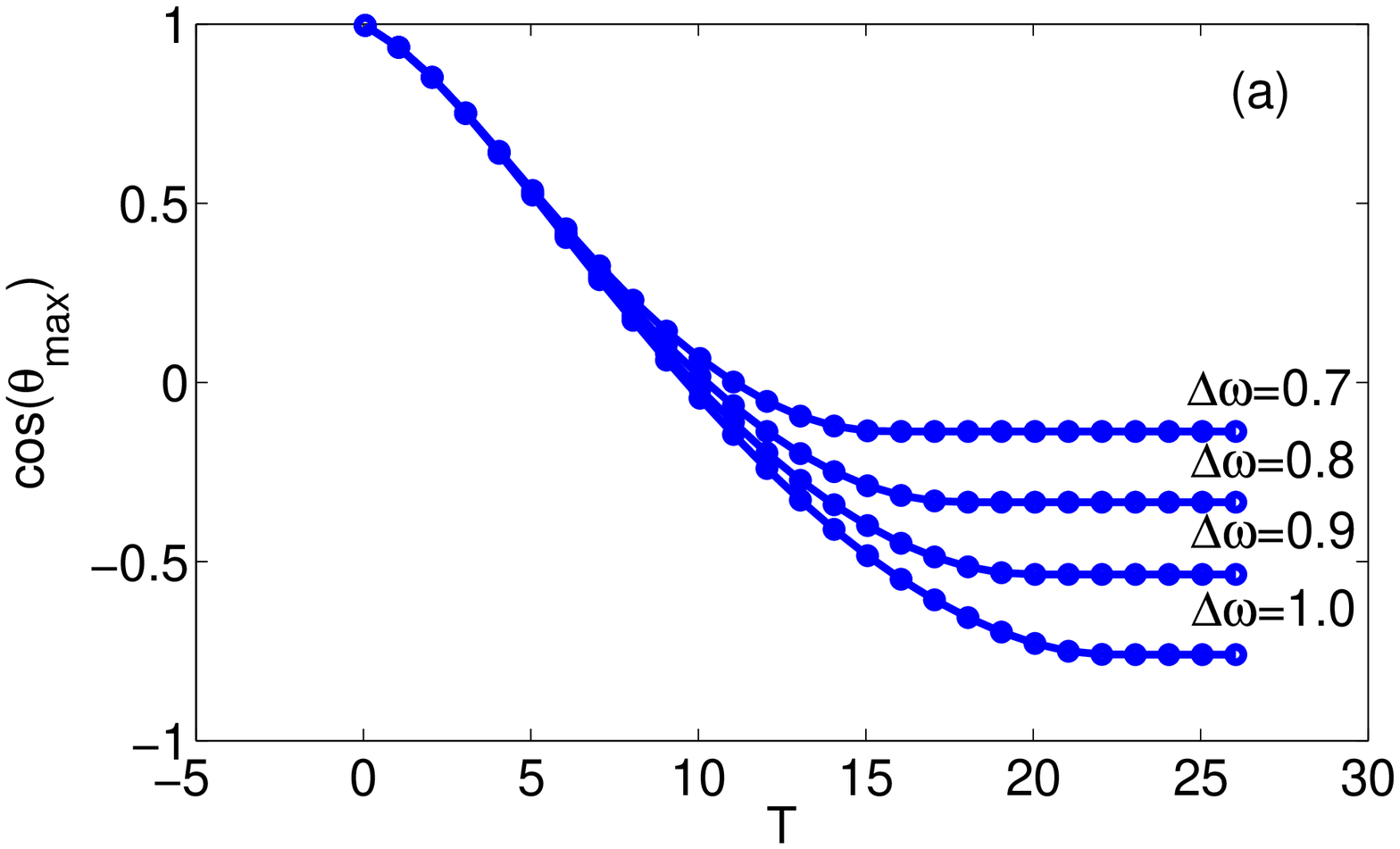}
    \label{Fig5a}
  \end{subfigure}
\begin{subfigure}[bp]{0.48\textwidth}
    \includegraphics[width=\textwidth]{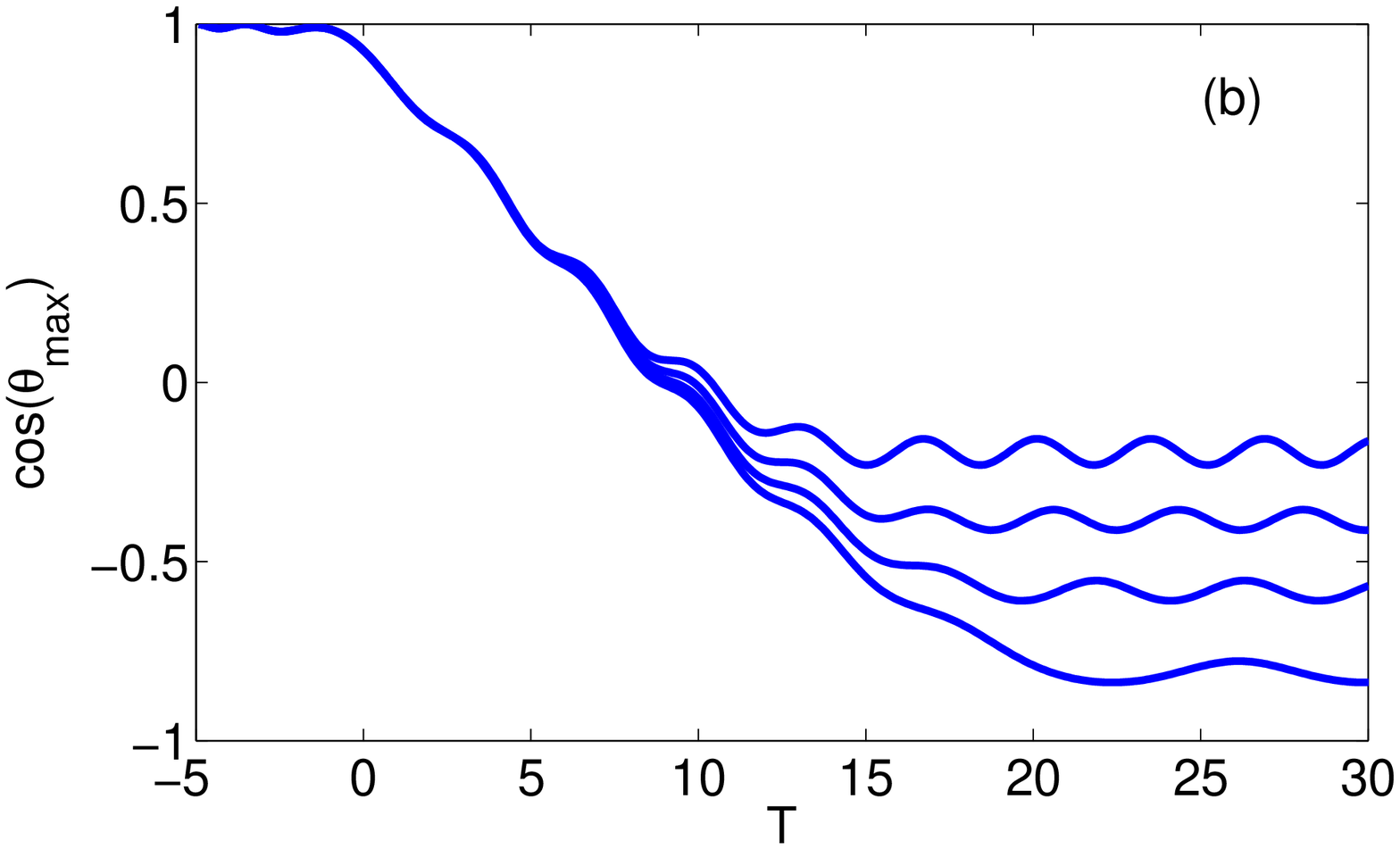}
    \label{Fig5b}
  \end{subfigure}
\caption{The axial magnetization component $m_{z}=cos(\protect\theta _{max})$
at the maximum $\protect\theta _{max}$ of the autoresonant excitation versus
slow time $T$. (a) the quasi-steady-state theory, (b) numerical LLG simulations.}
\label{Fig5}
\end{figure}
As an example, we have solved this problem numerically (for details of this
calculations see Appendix) and found $A$ for different values of $\Omega
_{d}^{\prime }=1+k^{2}-D$ in the case of $l=8$ ($k=0.393$) and $0.7\leqslant
D\leqslant 1.05$ (recall that $D\ $represents the driving frequency
deviation $D=\Delta \omega \sin (\alpha \tau /\Delta \omega )$ from the
linear resonance frequency in all our numerical simulations). Note that $%
\lambda \rightarrow \infty $ at the limiting values $A=0$ and $A=V_{\max }$
and, therefore, $\lambda $\ has a minimum at some $A$ between $0$ and $%
V_{\max }$. Therefore, for having the solution of equation $\lambda =l$, $l$
must be above this minimum. We have found that such solutions in our example
exists for $D<1.05$. \ The knowledge of $A$ allows to calculate the spatial
profile $\theta (\xi )$ by solving Eq. (\ref{S5_1}) in quadratures, but, for
simplicity, we have limited the calculation to just finding the maximum of
the autoresonant excitation $\theta _{\max }=\theta (l/2)$ by solving $%
V(\theta _{\max })=A$ at each time. We summarize this quasi-steady-state
evolution in Fig. 5a, showing $m_{z}(l/2)=\cos \theta _{\max }$ (circles)
versus slow time $T$ for four values of $\Delta \omega =0.7,0.8,0.9,$ and $%
1.0$. The results are in a good agreement with the full numerical LLG
simulations shown in Fig. 5b for the same $\Delta \omega $, confirming the
quasi-steady-state theory. Note that the quasi-steady-state calculations
involve solution of algebraic equations only, which is significantly simpler
that the full numerical simulations.

It is interesting to compare the amplitudes of the solutions in Fig. 5a to
those of the exact solitons for easy-axis magnets \cite{Kosevich}:%
\begin{equation}
\tan ^{2}(\theta _{\max }/2)=\Omega _{d}^{\prime }/(\Omega _{d}^{\prime
}-\Delta \omega )-1.  \label{Amp}
\end{equation}%
\begin{figure}[tp]
\centering\includegraphics[width=8.8cm]{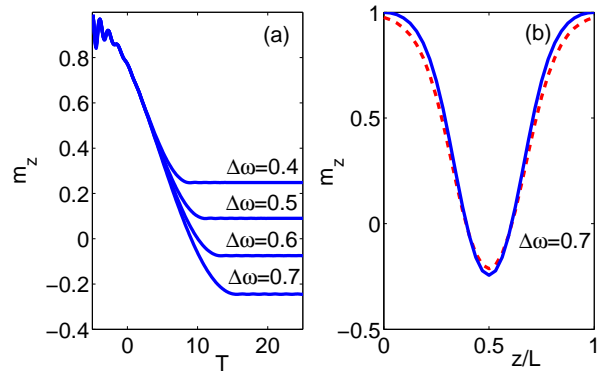}
\caption{The autoresonant excitation in the presence of dissipation $\protect%
\eta =0.01$ (LLG simulations). (a) The axial magnetization component $m_{z}=cos(\protect\theta %
_{max})$ at the maximum $\protect\theta _{max}$ of the autoresonant
excitation versus slow time $T$, (b) The comparison of the excited waveform
(solid line) with the exact dissipationless soliton solution \protect\cite%
{Kosevich} (dashed line).}
\label{Fig6}
\end{figure}
This equation yields $\cos (\theta _{\max })=-0.20,-0.38,-0.56,$ $-0.73$ for
$\Delta \omega =0.7,0.8,0.9,1.0$, respectively, in a very good agreement
with the values in Fig. 5a at the final time. It is also interesting to
discuss the effect of dissipation. Figure 6a shows the results of the LLG
simulations similar to those in Fig. 6b, but for $\eta =0.01$ and $\Delta
\omega =0.4,0.5,0.6,0.7$. The final amplitudes $m_{z}(l/2)=\cos \theta
_{\max }$ of the solutions in this cases were $0.31,0.13,-0.04$ and $-0.21$.
In addition, Fig. 6b shows the solution $m_{z}(z/L)$ found in numerical LLG
simulations (solid blue line) at $\Delta \omega =0.7$ and compares it to the
form of the exact soliton solution (red dashed line) \cite{Kosevich}. The
good agreement in the figure shows that the boundary conditions imposed
sufficiently far from the localized magnetization structure only slightly
deform its shape preserving the proximity to the soliton solution, i.e., the
autoresonant drive efficiently compensates for dissipation.
\begin{figure}[bp]
\centering\includegraphics[width=8.8cm]{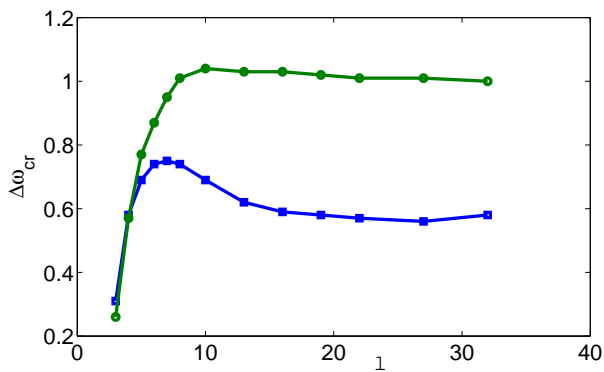}
\caption{The dependence of the critical driving frequency shift $\Delta
\protect\omega _{cr}$ on $l$: $\protect\eta =0$ (upper line) and $\protect%
\eta =0.01$ (lower line) for $\protect\varepsilon =2\protect\varepsilon _{th}
$ and $\protect\alpha =0.005$.}
\label{Fig7}
\end{figure}

The amplitude of the autoresonant steady-state solutions depends on the
final driving frequency shift $\Delta \omega $. We have found numerically
that there exists some critical value $\Delta \omega _{cr}$ such that stable
phase-locked solutions can be excited for $\Delta \omega <\Delta \omega _{cr}
$ only. One finds that $\Delta \omega _{cr}$ depends mainly on length $l$
and weakly on $\varepsilon $ and $\alpha $ if $\varepsilon >\varepsilon _{th}
$, see Fig. 7. In the dissipationless case, $\Delta \omega _{cr}$ is close
to unity for $l>7$ and rapidly decreases for $l<7$. Note that $l\approx 7$
is near the typical size $\Delta \xi $ of the solitons [see Eq. (\ref{Width}%
)], while the maximum value of $\Delta \omega _{cr}\approx 1.05$ is in
agreement with $D=1.05$ case in Fig. 9 in the Appendix. In the dissipative
case, the autoresonant driving compensates the effect of damping, providing
stability of the driven phase-locked soliton. In this case, $\Delta \omega
_{cr}$ has the maximum value of $0.75$ at $l\approx 7$, yielding the largest
amplitude of the autoresonant excitation for this damping parameter. We have
also found that with dissipation, $\Delta \omega _{cr}$ rapidly decreases
for $l<7$ and that stable solitons do not exist for $l<3$ with and without
dissipation.
\begin{figure}[tp]
\centering\includegraphics[width=8.8cm]{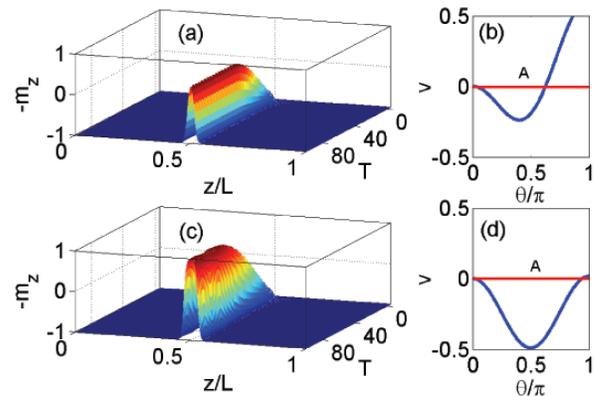}
\caption{The excitation of narrow magnetization structures using increased anisotropy of the
end sections of the particle to preserve $m_{z}=1$ boundary conditions. Panels (a)
and (c) show the $-m_{z}$ for $\Delta \protect\omega =0.7$ and $1.0$
respectively, while panels (b) and (d) present the corresponding effective
potentials $V$ and the quasi-energies $A$ (red horizontal lines) at final
time $T=100$.}
\label{Fig8}
\end{figure}

We conclude this Section by discussing the realization of the assumed
boundary conditions, i.e., that the spins of the nanoparticle are 'pinned'
at the boundaries ($m_{z}=1)$. It turns out that this condition can be
realized for free spins as well. Indeed, we can exploit the fact that the
proposed excitation approach involves a resonant driving mechanism.
Therefore, if one destroys the resonance at the ends of the nanoparticle,
the magnetization of the ends will not be affected by the resonant drive
and, thus, remain in the direction of the guiding magnetic field $\mathbf{H}$
as set initially. For example, if the anisotropy of the end sections of the
particle is larger than of its middle part, the ends will remain steady when
the driving field is tuned to resonate with the middle section. We
illustrate this approach in numerical LLG simulations in Figs. 8a and 8c showing two narrow structures of
different height excited in a nanoparticle of length $l=100$ with the two
end sections of length $l/3$ each having (normalized) anisotropy constant of
$1.25$, while the anisotropy constant of the middle $l/3$ section is $1.0$,
as before. In addition, Figs. 8b and 8d show the effective potential for
both excitations and the corresponding quasi-energies, $A\approx 0$, as
expected for solitary waves. In these simulations, we set $m_{z}=1$ at $\xi
=0,l$ initially, but did not impose this condition at later times. The
figure shows two solutions of different amplitude excited along the particle
for $\Delta \omega =0.7$ and $1.0.$ The rest of the parameters were $\alpha
=0.005$, $h_{0}=5$, $\eta =0.005$, $\varepsilon =1.2\varepsilon _{th}$. As
expected, the boundary condition of $m_{z}=1$ at the ends of the particle is
preserved (the ends are off resonance) and the excited autoresonant
solutions have a width $\Delta \xi \approx 7$ [see Eq. (\ref{Width})] much
smaller than the length of the particle.

\section{Conclusions}

In conclusion, we have suggested and analyzed the autoresonant approach to
excitation of LLG localized solutions in easy axis ferromagnetic
nanoparticles. The approach involved driving the system under fixed boundary
conditions by a spatially uniform, but oscillating magnetic field having
chirped frequency, which passes through the linear resonance with initially
uniform magnetization. When the driving amplitude exceeded a threshold $%
\varepsilon _{th}$, we have observed formation of large amplitude rotating
magnetization structures (see Fig. 1). We have used a weakly nonlinear
Lagrangian formulation in the dissipationless limit of the model for
calculating the threshold [Eq. (\ref{S3_11})] and discussed the effect of
dissipation on the threshold [Eq. (\ref{S4_7})]. All these predictions were
in a good agreement with simulations. We have also analyzed fully nonlinear
autoresonant quasi-steady-state localized solutions using a simple analog of
a quasi-particle in an effective potential (see developments in Sec. V). In
these calculations, the decreasing driving frequency gradually approached
some fixed target value at distance $\Delta \omega $ from the linear
resonance, which also defined the target amplitude of the excited solution.
We found numerically that there exists a critical $\Delta \omega _{cr}$,
such that the excited structures remained stable if $\Delta \omega <\Delta
\omega _{cr}$. This critical $\Delta \omega _{cr}$ depended mainly on the
dimensionless length $l$ and damping parameter $\eta $ and decreased rapidly
for $l<7$ (the approximate width of the solution). In the dissipationless
case, $\Delta \omega _{cr}\approx 1.05,$ yielding almost complete inversion
of magnetization at the solution maximum. We have found that the
autoresonant drive effectively compensates the effect of dissipation on the
excited solutions, but the dissipation lowers their amplitude. Finally, we
have suggested and illustrated an approach to realization of the assumed
fixed boundary conditions by increasing the anisotropy of the end sections
of the particle. A further development of the Whitham's-type variational
theory in order to explain the stability (seen in simulations) of the
chirped-driven large amplitude localized solutions is important. Finally, it
is known that the dissipationless LLG equation in 1D is integrable and has a
large variety of solutions \cite{Kosevich}. The autoresonant excitation and
control of some of these solutions comprises another important goal for
future research.

\section{Acknowledgement}

This work was supported by the Israel Science Foundation Grant No. 30/14 and
the Russian state program AAAA-A18-118020190095-4.

\section{Appendix}

\begin{figure}[tp]
\centering\includegraphics[width=8.8cm]{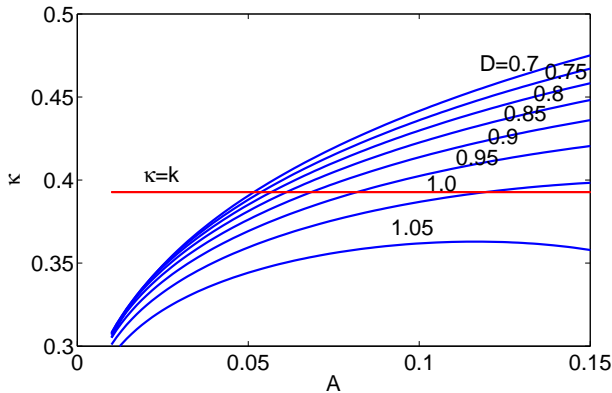}
\caption{The parameter $\protect\kappa=\protect\pi/\protect\lambda $ versus
the quasi-energy $A$ for different driving frequency shifts $D$ and $l=8$.
Condition $\protect\kappa =k$ can be satisfied for $D<1.05$ only.}
\label{Fig9}
\end{figure}
In calculating the quasi-energy $A$ of the slow evolution of the
autoresonant quasi-steady-state, instead of solving Eq. (\ref{Lambda})
defining $\lambda $, we proceed by introducing the action
\begin{equation}
I(A)=\frac{2}{\pi }\int_{0}^{\pi }Re[\sqrt{2(A-V)}]d\theta .  \label{S5_3}
\end{equation}%
Then, since
\begin{equation}
\kappa =\frac{\pi }{\lambda }=\left( \frac{dI}{dA}\right) ^{-1},
\label{S5_3a}
\end{equation}
we calculate the action for different values of $\Omega _{d}^{\prime
}=1+k^{2}-D$ and take its derivative $(dI/dA)^{-1}$ numerically to find $%
\kappa (A)$. The results are shown in Fig. 9 in the case of $l=8$ ($k=0.393$%
) and $0.7\leqslant D\leqslant 1.05$. Such calculations and using $\Omega
_{d}^{\prime }$ of form (\ref{S3_om}) allow to find the quasi-energy $%
A $ such that $\kappa =k$ ($\lambda =l$) at different times. Note that $%
\kappa =0$ at the limiting values of $A=0$ and $V_{\max }$ and, therefore, $%
\kappa $\ has a maximum at some $A$ between $0$ and $V_{\max }$, as can be
seen in Fig. 5 for $D=1.05$. Thus, for having the solution with $\kappa =k$
in the driven problem ($k=0.393$ in our example), $k$ must be below this
maximum as for $D\leqslant 1.0$ in Fig. 5, while no such solution exists for
$D=1.05$.

\end{document}